\documentclass[pre,aps,preprint,showpacs]{revtex4} 
\usepackage{revsymb,amsmath}
\usepackage{graphicx,psfrag}
\usepackage{nicefrac}

\begin{document}
\author{Michael Schulz}
\affiliation{Abteilung Theoretische Physik, Universit\"at Ulm\\D-89069 Ulm Germany}
\email{michael.schulz@uni-ulm.de}
\author{Steffen Trimper, Knud Zabrocki}
\affiliation{Fachbereich Physik, Martin-Luther-Universit\"at,D-06099 Halle Germany}
\email{trimper@physik.uni-halle.de}
\title{Spatiotemporal Memory in a Diffusion-Reaction System}
\date{\today }
\begin{abstract}
We consider a reaction-diffusion process with retardation. The particles, immersed in traps initially, 
remain inactive until another particle is 
annihilated spontaneously with a rate $\lambda$ at a certain point $\vec x$. In that case 
the traps within a sphere of radius $R(t)= v t^{\alpha}$ around $\vec x$ will be activated 
and a particle is released with a rate $\mu$. Due to the competition between both 
reactions the system evolves three different time regimes. While in the initial time 
interval the diffusive process dominates the behavior of the system, there appears a 
transient regime, where the system shows a driveling wave solution which tends to a 
non-trivial stationary solution for $v \to 0$. In that regime one observes a very slow 
decay of the concentration. In the final long time regime a crossover to an exponentially 
decaying process is observed. In case of $\lambda = \mu$ the concentration is a conserved 
quantity whereas for $\mu > \lambda$  the total particle number tends to zero after a 
finite time. The mean square displacement offers an anomalous diffusive behavior where the dynamic 
exponent is determined by the exponent $\alpha $. In one dimension the model can be solved 
exactly. In higher dimension we find approximative analytical results in very good agreement 
with numerical solutions. The situation could be applied for the development of a bacterial 
colony or a gene-pool.

\end{abstract}
\maketitle

\section{Introduction}
A broad variety of different problems in physics and biology can be formulated in terms of 
generalized evolution equations including delay and feedback effects \cite{mur,fra}. Such 
memory-controlled effects should be a further unifying feature of complex physical \cite{6a,6b} 
as well as biological systems \cite{6c} far from equilibrium. Recently \cite{myh} memory effects 
have been qualified as the key feature in describing dynamical systems. Distributed delays are 
able to stabilize ecological systems \cite{etf}. Even the traffic flow can be characterized by the scaling 
properties of the underlying memory function \cite{he,sh}. The whole history of systems offering 
self-organized criticality as earthquakes is discussed in \cite{lag}. Whereas most of the papers 
are addressed to a purely time dependent but homogeneous memory, the present one gives an 
extension to spatiotemporal processes. Based on our recent studies of several evolution models 
\cite{bst,st1,zab,zab1}, where such non-Markovian memory effects had been enclosed, and on the 
analysis of chemical reactions with a long range memory \cite{trizab}, the present study is focused to 
a diffusion-reaction behavior with short and long range memory couplings. As an essential new 
element we include the case that retardation effects are relevant. Thus cells are able to interact 
as well as by short-range forces such as adhesion and through either long-range forces or retarded 
interaction via chemical signals \cite{gri}. These effects have an essential influence on the long 
time behavior, but they have also an impact on the dynamics in an intermediate time regime. As the 
consequence of the interplay between conventional diffusive transport and annihilation and creation 
processes with feedback we demonstrate the system evolves slowly varying plateau in the concentration.\\ 
\noindent Our analysis can be grouped into the continuing interest in feedback couplings. A memory dominated 
behavior is well established in analyzing the freezing processes in under-cooled liquids \cite{l,goe}, where 
the underlying mathematical representation is based on a projector formalism proposed by Mori \cite{mori}. 
As the result of the projection procedure the irrelevant degrees of freedom, which are rapid fluctuating ones, 
contribute to the time evolution of the relevant degrees of freedom as well as by instantaneous and by 
delay-controlled terms. Since the projection formalism is not restricted to selected systems, the 
modification of the evolution equation due to memory effects is rather generic. The relevant degrees of 
freedom offering a slowly varying behavior could be the concentration of certain species or the 
probability density for the particles. The crucial influence of memory effects can be illustrated by 
considering a single particle moving in a disordered environment. Due to the strong disorder each 
member of an equally prepared ensemble makes experience by its own surrounding, which is even modified 
by the random walk of the single particle itself. Such a behavior can be described by a non-linear 
Fokker-Planck equation with an additional memory term as had been demonstrated in \cite{ss}. 
The analytical results, based on a one-loop renormalization group analysis, is in 
accordance with numerical simulations \cite{bst} and further analytical studies \cite{st1}. Whereas in that 
approach the memory effects are believed to be originated by the inherent non-linear interaction of the 
many particle system themselves, i.e. the time scale of the memory is determined by the relevant 
variable itself, there is a broad class of models which are subjected to external delay effects \cite{o,g,fe}; 
for a survey and applications in biology, see \cite{mathbio}.\\ 
Recently \cite{fk} memory effects in correlated anisotropic diffusion are studied in nanoporous crystalline 
solids. Likewise the effects of transport memory are discussed within the Fisher`s equation \cite{ah}, 
ratchet devises \cite{sr} also applicable for bacterial population dynamics \cite{kk}. There appear a non-linear damping and 
traveling wave solutions \cite{abk}. The transport with memory, depending on the survivability of a 
population, is analyzed in \cite{h}.\\ 
In case the transport is realized after a spatiotemporal accumulation process the time evolution 
of the probability or the concentration of a different species could be also dependent on the history 
of the sample to which it belongs. Thus the evolution of the relevant variable $p(\vec x, t)$ has to be 
supplemented by memory terms. One possible but rather general structure of an evolution equation 
with memory should read \cite{spie}  
\begin{equation}
\partial _{t}p(\vec x, t) = \mathcal{M}(\vec x,t; p, \nabla p)  
+\int\limits_{0}^{t}dt\,'\int\limits_{-\infty }^{\infty }d^dx^{\prime}
\mathcal{K}(\vec x- \vec x^{\prime},t-t^{\prime}; p, \nabla p)\mathcal{L}(\vec x^{\prime}, t^{\prime}; p,\nabla p)\,.
\label{2}
\end{equation}
This equation is of convolution type. Whereas the operator $\mathcal{M}$ characterizes the instantaneous and 
local processes, both operators $\mathcal{K}$ and $\mathcal{L}$ are responsible for the delayed processes. In 
general the operators $\mathcal{M}, \mathcal{K}$ and $\mathcal{L}$ may be also non-linear in $p(\vec x, t)$ and 
$\nabla p(\vec x,t)$. They have to be specified according to the physical situation in mind. In particular 
we have shown \cite{preis,spie} that the form of the operator $\mathcal{L}$ is restricted when $p(\vec x,t)$ is 
assumed to be conserved. Regarding the large variety of systems with feedback couplings it seems to be worth 
to study simple models, which still include the dynamical features of evolution models as conservation of 
the relevant quantity $p(\vec x, t)$ and moreover, a time-delayed coupling.\\
In the present paper we extend the analysis by considering another kind of memory not included in 
Eq.~(\ref{2}). To that aim let us consider a diffusion-reaction model with retardation for a concentration 
$p (\vec x, t)$. Different to Eq.~(\ref{2}) there appears a spatiotemporal coupling well known from 
retarded potentials. Each particle makes a random walk described by a diffusion equation \cite{ah} 
in the continuum limit. Moreover, it can be annihilated with the rate $\lambda $ spontaneously. As a 
competing process a self-organized creation of particles with rate $\mu $ is introduced. However the 
additional process is subjected to a retardation. The situation corresponds to a diffusion-reaction 
process where the particles are able to emit permanently signals. After a finite running time inactive 
particles will be excited.         

\section{Different realizations and the general solution }
To be specific let us consider a diffusion-reaction model for the concentration $p (\vec x, t)$ of the form
\begin{eqnarray}
\frac{\partial p(\vec x, t)}{\partial t} &=& D \nabla ^2 p(\vec x, t) - \lambda p(\vec x, t) \nonumber\\
&+& \mu \int_{-\infty}^{\infty} d^dx^{\prime}\, \Phi \left(\vec x\,^{\prime},\, t - 
\frac{\mid \vec x - \vec x\,^{\prime} \mid}{v} \right)\, 
N(\mid \vec x - \vec x\,^{\prime} \mid) \Theta \left(t - \frac{\mid \vec x - \vec x\,^{\prime} \mid}{v} \right) 
\label{eq1}
\end{eqnarray}
Particles, performing a random walk with diffusion constant $D$ are annihilated with the rate $\lambda $ 
at a certain point $\vec x$ at time $t$. The subsequent term describes the creation of particles with 
the rate $\mu $. This creation process is triggered by a signal that will be transmitted by each particle 
at the time $t^{\prime} =  t - \frac{\mid \vec x - \vec x\,^{\prime} \mid}{v} $. With other words, 
particles are created whenever the signal wave, originated from any point $\vec x\,^{\prime}$ at time 
$t^{\prime}$ is reached the point with the coordinate $\vec x$ at the observation time $ t \leq t^{\prime}$. 
The quantity $\Phi (\vec x\,^{\prime}, t^{\prime})$ is determined by the concentration $p$, and different 
realizations will be discussed below. To complete the model we have to fix the normalization factor 
$N(r)$ by $\int r^{d-1} N(r) d\Omega_d = 1$ which leads to 
$$
N(R) = \frac{1}{\Omega _d R^{d-1}}\quad \rm{with} \quad \Omega _d = \frac{2 \pi ^{d/2}}{\Gamma (d/2)}\,.
$$
After Fourier transformation Eq.~(\ref{eq1}) reads
$$
\frac{\partial p(\vec k, t)}{\partial t} + [ D \vec k^2 + \lambda ] p(\vec k, t) = 
\mu \Gamma (d/2) 2^{d/2-1} \int _0^t dt^{\prime} [k v(t -t^{\prime}]^{1-d/2} J_{d/2-1}(k v(t -t^{\prime}) 
\Phi (\vec k, t^{\prime})\,,
$$
where $J_a(x)$ is the Bessel function. Performing Laplace transformation with respect to time we get 
\begin{equation}
(z + D k^2 + \lambda ) p(\vec k, z) = p(\vec k, t=0) + 
\frac{\mu v}{z} F\left(\frac{1}{2}, 1, \frac{d}{2}; - \frac{k^2 v^2}{z^2}\right) \Phi (\vec k, z)\,.
\label{hyp}
\end{equation}
Here, F(a,b,c; y) is the hypergeometric function. For the further discussion the function 
$\Phi (\vec x, t) $ has to be specified. Here we discuss three cases
\begin{eqnarray}
(i) \quad \Phi (\vec x, t) &=& p(\vec x,0) \delta (t) \equiv p_0(\vec x) \delta (t) \nonumber\\
(ii) \quad \Phi (\vec x, t) &=& p (\vec x, t)\nonumber\\
(iii) \quad \Phi (\vec x, t) &=& \frac{\partial p (\vec x, t)}{\partial t}\,.
\label{special}
\end{eqnarray}
In case (ii) the concentration is given by 
\begin{equation}
p(\vec k, z) = \frac{p_0(\vec k)}
{z + D k^2 + \lambda + \frac{\mu v}{z} F(\frac{1}{2},1,\frac{d}{2}; - \frac{k^2v^2}{z^2})}\,.
\label{cii}
\end{equation}
In case (iii) the retardation is manifested by a coupling to the time variation of the 
concentration itself. It results
\begin{equation}
p(\vec k, z) = \frac{p_0(\vec k)\left[ 1 - \frac{\mu v}{z} F(\frac{1}{2},1,\frac{d}{2}; - \frac{k^2v^2}{z^2})\right]}     
{z + D k^2 + \lambda + \frac{\mu v}{z} F(\frac{1}{2},1,\frac{d}{2}; - \frac{k^2v^2}{z^2})}\,.
\label{ciii}
\end{equation}
Both equations do not give rise to an transient regime as it will be observed in the first case. 
Therefore we restrict the further discussion mainly to the first case which is discussed in detail 
in the forthcoming section.  

\section{Long range memory}
In case of 
\begin{equation}
\Phi (\vec x, t) = p_0(\vec x) \delta (t)
\end{equation}
the model offers additional specifications like conservation and anomalous behavior. 
The realization given by the last equation means that the creation process is triggered by a signal 
that had been transmitted at the initial time $t = 0$, where the particles are distributed with the 
concentration $p_0(\vec x)$. The signal is presumed to be permanent existent in the whole space like 
the microwave background radiation in cosmology. The activation of particles by this background signal 
can be further generalized in the following manner. If at an arbitrary point $\vec x$ 
a particle is annihilated then the activation process sets in. However, to simulate the influence 
of the backward signal it should be restricted to a sphere with radius 
\begin{equation}
R(t) = v t^{\alpha }
\label{kug}
\end{equation}
around the mentioned point at $\vec x$, where the exponent will be specified below. 
Only in case $\alpha = 1$ the quantity $v$ can be interpreted as a real velocity. 
While the particles perform a random walk, they will 
be annihilated with a certain rate $\lambda $ and created with the rate $\mu $. This process is governed 
by signals coming from the a sphere with radius $R(t)$ around the point with the coordinate $\vec x$. 
The situation in mind is simply described by the following evolution equation
\begin{equation}
\frac{\partial p(\vec x, t)}{\partial t} = D \nabla ^2 p(\vec x, t) - \int d^dr \left[\,\lambda p(\vec x, t)
- \mu\,N(r) p(\,\vec x - \vec r,0)\right] \delta (r - R(t))\quad \mbox{with}\quad r = \mid \vec r\mid .
\label{eq1a}
\end{equation}
Here the first term describes conventional diffusion with the diffusion constant $D$. 
The second term stands for the annihilation of particles with rate $\lambda $. The last terms 
models the situation discussed before. Particles are created with a rate $\mu $ whenever the signal wave 
originated from the initial state of the system is reached the spatial point $\vec x$. This process 
corresponds to a kind of self-stabilizing of the system, especially for $\lambda = \mu$, where 
the total particle number is conserved. 
If both rates are different one obtains for the total number of particles $P(t)$ the relation
\begin{equation}
P(t) = P(0) \left [ \frac{\mu }{\lambda } + e^{-\lambda t}\,\left(1 - \frac{\mu }{\lambda }\right) \,\right]\,,
\label{eq1b}
\end{equation}
which is also independent of the realization of the function $R(t)$. In the limiting case 
$\lambda = \mu$ the conservation is guaranteed. In case of a vanishing rate $\lambda = 0$ 
there occurs a linear increase of the concentration according to $P(t) = P(0)(1 +\mu t)$. 
Let us remark that in the opposite case 
where growth and death rate will be interchanged, i.e. $\lambda \to - \lambda $ and $\mu \to - \mu $ 
the last equation leads to a complete extinction of the species after a finite time 
$$
t_e = \frac{1}{\lambda } \ln \left(\frac{\mu }{\mu - \lambda }\right)\,,
$$
provided $\mu > \lambda $. As one can see from Eq.~(\ref{eq1a}) there appears a driving force between the 
concentration of the spatial point $\vec x$ and a point far from that at $\vec x - \vec R(t)$, 
i.e. all spatial points immersed in a sphere of radius $R(t) = vt^{\alpha} $ contribute to the instantaneous 
development of the system. The model describes a situation, where a set of particles performing a 
random walk, carry the concentration in the past like a jet. In case of zero parameter $v = 0$ the 
system reveals an extreme long range memory, characterized by a direct coupling of the probability 
$p(\vec x, t)$ to its initial value $p(\vec x, 0)$. Due to this long range coupling there exits 
a non-trivial stationary solution as demonstrated in \cite{zabrocki}. This stationary solution 
is originated by the balance of the diffusive current and the memory induced current proportional 
$p(\vec x, t) - p(\vec x, 0)$.\\
In the present approach we consider a new situation consisting of the existence of an active local 
environment of radius $R(t)$ around each spatial point at $\vec x$. The situation corresponds 
to the presence of traps in which particles are immersed. These traps will be installed at initial time 
$t=0$. However the barrier to overcome is to high, so that the particles are confined within the traps.
The particle inside a trap remains inactive until another particle is annihilated spontaneously 
at a spatial point $\vec x$. Let us assume that through the annihilation process energy will be released 
so that the particles are able to overcome the barrier of the traps and the particle included become 
active with a rate $\mu$ provided they are in a sphere of radius $R(t)$ 
around $\vec x$. With other word, instead of a long range memory there is a short range one induced 
by a coupling of the current concentration $p(\vec x, t)$ to the value of the distribution 
$p(\vec x - \vec R(t), 0)$. In that case we demonstrate that the stationary solution 
disappears and instead of that the system evolves a very slow decay in a broad transient time regime.  
In the limiting case of vanishing $v$ the stationary solution is restored. The transient regime is 
likewise originated due to the balance of the diffusive current and the additional 
memory current proportional to $p(\vec x, t) - p(\vec x - \vec R(t))$.  
Assuming the quantity $p(\vec x, t)$ is proportional to the probability to find the particle at the 
spatial point $\vec x$ at time $t$ we get for the mean square displacement $s(t) = \langle \vec x^2 \rangle$ the 
evolution equation 
\begin{equation}
\frac{d s(t)}{dt} = 2 d D - \lambda s(t) + \mu [ s(0) + R^2(t)]\,.
\label{ms}
\end{equation}
In case of $R(t) = v t^{\alpha }$ the last equation can be solved resulting in
\begin{equation}
s(t) = \frac{2 d D}{\lambda }\left(1 -e^{-\lambda t}\right) + \mu  v^2 (- \lambda )^{1+2 \alpha }\,
\gamma (1+ 2 \alpha , - \lambda t)
\label{ms1}
\end{equation}
Here $\gamma (a, y)$ is the incomplete gamma function. For nonzero $\lambda $ the asymptotic behavior 
of the displacement is for $\lambda t \gg 1$ given by  
$$
s(t) \simeq \frac{t^{2\alpha }}{\lambda }\,, 
$$
while in the case of vanishing annihilation rate $\lambda = 0$ one gets exactly
$$
\lim_{\lambda \to 0} s(t) = 2 d D t + \frac{\mu  v^2}{1 + 2 \alpha} t^{1 + 2 \alpha } \,.
$$
The growth term in Eq.~(\ref{eq1}) gives rise to anomalous diffusion which dominates for $\alpha > 0$ 
the long time behavior of the system. In particular, even for $\alpha =1/2$ globally the system 
offers a ballistic behavior. In case of conservation of $p(\vec x, t)$ the last relation leads 
to diffusion due to $\lambda = \mu$.

\section{Solution}
Since the system exhibits no general analytical solution for an arbitrary behavior of $R(t)$ we 
restrict the further discussion to the case $\alpha = 1$. Fourier transformation of Eq.~(\ref{eq1}) yields 
\begin{equation}
\partial_t p(\vec k, t) = - (D k^2 + \lambda ) p(\vec k, t)  + \lambda  A p_0(\vec k) 
\frac{J_{\frac{d-2}{2}}(kR(t))}{[kR(t)]^{\frac{d-2}{2}}}\quad\text{with}\quad A = \Gamma (d/2)2^{d/2-1}; \,\,
k = \mid \vec k \mid\,.
\label{eq3}
\end{equation}
Here $J_{\nu} (y) $ is the Bessel function. Although the last equation is only valid for dimension $d > 1$ 
\cite{ryshik}, a separate transformation reveals that the one dimensional case with    
\begin{equation}
\partial_t p(k, t) = - (D k^2 + \lambda ) p(k, t)  + \mu p_0(k) \cos(k R(t))\,,
\label{eq4} 
\end{equation}
is also included.
To analyze whether the model offers a nontrivial stationary solution it is appropriate to consider the  
Laplace transformation $p(\vec k, z)$. Aside of special cases the transformation can be fulfilled only 
for $R(t)= v t$, i.e. for $\alpha = 1$. From Eq.~(\ref{eq3}) we find
\begin{equation}
p(\vec k, z) = p_0(\vec k) \frac{1 + \nicefrac{\mu }{z} F(\frac{1}{2},1,\frac{d}{2}; - \frac{k^2 v^2}{z^2})}
{z + Dk^2 + \lambda }\,,
\label{eq5}
\end{equation}
where $F(a,b,c;y)$ is the hypergeometric function. The result is always valid for $d \geq 1$.
The long time limit results for $z \to 0$.  Because there is no general asymptotic behavior of 
the hypergeometric function $F(a,b,c,y)$ for $y \to \infty$ and arbitrary dimension $d$ 
we will discuss different dimensions separately.  

\subsection{Exact solution for $d=1$}
In the one- dimensional case the solution for $p(x,t)$ can be expressed by elementary functions. 
In $d=1$ the evolution equation reads
\begin{equation}
\frac{\partial p(x,t)}{\partial t} = D \frac{\partial^2 p(x,t)}{\partial x^2} - \lambda p(x,t) + 
\frac{\mu}{2}  [p(x-R(t),0) - p(x+R(t),0)] \quad\text{with}\quad p(x,0) = p_0(x)\,,
\label{d11}
\end{equation}
where $R(t) = v t$ is defined by Eq.~(\ref{kug}) with the exponent $\alpha = 1$. Performing Fourier 
transformation with respect to the spatial coordinate we can solve the the resulting equation 
for  $p(k,t)$. According to Eq.~(\ref{eq4}) we get
\begin{equation}
p(k,t)= p_0(k) e^{-(Dk^2 + \lambda )t} + 
\frac{p_0(k)\mu}{2} \left(\frac{ e^{ikvt} - e^{-(Dk^2 + \lambda )t}}{Dk^2 + ikv + \lambda } + c.c.\right)\,.
\label{d12}
\end{equation}
By solving this equation one observes that the three terms describe three different time regimes. 
Whereas the first one, as solution of the homogeneous equation, represents the initial time interval, the second 
term gives rise to a traveling wave behavior which tends for $v=0$ to a space dependent 
stationary solution. The last term is responsible for the final time interval and leads to an 
exponential decay. After a lengthy but straightforward backward transformation we find 
$p(x,t) = p_i(x,t)+ p_s(x,t) + p_f(x,t)$ with
\begin{eqnarray}
p_i(x,t) &=& \frac{e^{-\lambda t}}{\sqrt{4\pi D t}}\int dx' p_0(x^{\prime}) e^{-\frac{(x-x^{\prime})^2}{4Dt}} 
\nonumber\\
&&\nonumber\\
p_s(x,t) &=& \frac{2 \mu }{\Omega} \int dx^{\prime} p_0(x^{\prime}) \left[\Theta (x-x^{\prime}+vt) 
e^{-\nu _1 (x - x^{\prime} + vt)} 
+ \Theta (x^{\prime} - x -  vt) e^{- \nu  _2 (x^{\prime}- x - vt)}\right] \nonumber\\
&&\nonumber\\
p_f(x,t) &=& -\frac{\mu  e^{-\lambda t}}{\Omega } \int dx^{\prime} p_0(x^{\prime}) 
\left\{e^{-\nu _1(x-x^{\prime}- D t \nu _1)}\text{erfc}\left(\frac{2\nu _1 D t - (x-x')}{\sqrt{4Dt}}\right)\right.
\nonumber\\
&+& \left. e^{\nu _2(x-x'+ D t \nu _2)} \text{erfc}\left(\frac{2 \nu _2 D t + (x-x')}{\sqrt{4Dt}}\right)\right\}\nonumber\\
&&\nonumber\\
\mbox{with}\quad \Omega &=& \sqrt{v^2 + 4 \lambda D},\quad \nu _{1/2} = \frac{\Omega \mp v}{2D} > 0\,. 
\label{d13}
\end{eqnarray}
In the limiting case $v = 0$ the part $p_s(x,t)$ tends to a non-trivial stationary solution 
$$
p_s(x) = \frac{\mu}{\sqrt{\lambda D}} \int_{-\infty}^{+\infty} dx' p_0(x') e^{-\frac{\mid x - x'\mid}{\xi }}\quad 
\text{with}\quad \xi = \sqrt{\frac{D}{\lambda}}\,. 
$$
This result is in accordance with a previous study \cite{zabrocki}. The behavior of $p(x,t)$ is controlled by 
a single dimensionless parameter 
$$
\kappa = \frac{v}{\sqrt{\lambda D}}\,,
$$
which is the ratio of the two velocities $v$ and $\sqrt{\lambda D}$.

\subsection{Solution in higher dimensions}
Guided by the solution in $d=1$ let us analyze the three dimensional case. In $d=3$  
the hypergeometric function in Eq.~(\ref{eq5}) can be expressed by an elementary function 
resulting in
\begin{equation} 
p(\vec k, z) = p_0(\vec k) \frac{1 + \nicefrac{\mu }{kv } \arctan (\frac{kv}{z})}{z + Dk^2 + \lambda }\,.
\label{eq6}
\end{equation}
Although we are not able to get a general solution for $p(\vec x,t)$, one can estimate the solution in different 
regimes. Denote $l$ a typical length scale and $\tau $ a characteristic time scale we can distinguish two cases:  
(i) $v \gg \frac{l}{\tau }$ and (ii) $v \ll \frac{l}{\tau }$. Due to $\frac{kv}{z} \gg 1$ in the first case the last 
equation leads to 
\begin{equation} 
p(\vec k, z) \approx p_0(\vec k) \frac{1 + \frac{\pi \lambda }{2kv }}{z + Dk^2 + \lambda }\,.
\label{eq7}
\end{equation}
From here we get an exponential decay $p(\vec k, t) \propto \exp(-[Dk^2 + \lambda ]t)$\,. In more detail we obtain
\begin{eqnarray}
p(\vec x,t) &\approx & p_i + p_f = \frac{e^{-\lambda t}}{(4\pi Dt)^{3/2}}\int d^3x^{\prime} p_0(\vec x^{\prime})
e^{-\frac{(\vec x - \vec x^{\prime})^2}{4Dt}} \nonumber\\
&+& \frac{\mu e^{-\lambda t}}{8\pi vDt}\int d^3x^{\prime} p_0(\vec x^{\prime})
e^{-\frac{(\vec x - \vec x^{\prime})^2}{4Dt}} 
{}_1F_1\left(1/2,3/2;\frac{(\vec x - \vec x^{\prime})^2}{4Dt}\right)\,.
\label{eq7a}
\end{eqnarray}
Here ${}_1F_1$ is the degenerate hypergeometric function which can be expressed by elementary functions: 
${}_1F_1 (1/2, 3/2, y) = \sqrt{\pi/4v}\, \text{erfi}(\sqrt{y})$ \cite{ryshik}. Notice that due to the condition 
$\frac{kv}{z} \gg 1$ the result can be not extrapolated to $v = 0$.  In the opposite case 
$\frac{kv}{z} \ll 1$ Eq.~(\ref{eq6}) is rewritten in the following form 
\begin{eqnarray}
p(\vec k, z) &\approx & \frac{p_s(\vec k)}{z} + \Psi (\vec k, z)\quad\mbox{with}\nonumber\\
p_s(\vec k) &=& \frac{p_0(\vec k) \mu }{Dk^2 + \lambda },\quad 
\Psi (\vec k,z)= p_0(\vec k)\frac{Dk^2 + \lambda - \mu }{(Dk^2+\lambda )(z+ Dk^2 + \lambda )}\,.
\label{eq7b}
\end{eqnarray}  
Due to the $1/z$ singularity in the last equation leads to  a\,(pseudo)stationary solution $p_s$. 
which reads after Fourier transformation
\begin{equation}
p_s(\vec x)= \frac{\mu }{4\pi D}\int \frac{p_0(\vec x\,')}{\mid \vec x - \vec x\,' \mid} 
\exp\left(-\frac{\mid \vec x - \vec x\,' \mid}{\xi }\right)\quad\mbox{with}\quad \xi = \sqrt{\frac{D}{\lambda }}\,.
\label{eq7c}
\end{equation}
Notice that $p_s$ is independent of $v$. This stationary solution is in accordance with \cite{zabrocki}. 
The correction to this stationary solution reads after Fourier transformation
\begin{eqnarray}
\Psi (\vec x,t) &=&  \frac{e^{-\lambda t}}{(4\pi Dt)^{3/2}}\int d^3x\,'p_0(\vec x\,') 
\exp(-(\vec x -\vec x\,')^2/4Dt))\nonumber\\
&+& \frac{\mu }{8 \pi  D} \int d^3x\,' \frac{p_0(\vec x^{\prime})}{\mid \vec x - \vec x^{\prime} \mid}
\left\{e^{\mid \vec x - \vec x^{\prime} \mid/\xi } 
\text{erfc}\,\left(\sqrt{\lambda t} + \frac{\mid \vec x - \vec x^{\prime} \mid}{\xi }\right)\right.\nonumber\\
&-& \left.e^{-\mid \vec x - \vec x^{\prime} \mid/\xi }\text{erfc}\,\left( \sqrt{\lambda t} - 
\frac{\mid \vec x - \vec x^{\prime} \mid}{\xi }\right)\right\}\,. 
\label{eq7d}
\end{eqnarray}
With other word we find a transient behavior in an intermediate time interval, whereas in the long time limit 
the concentration decays to zero. Such a behavior is similar to the $\alpha$ - relaxation in undercooled liquids.\\
For $d=2$ the hypergeometric function can be expressed by an elementary function. From Eq.~(\ref{eq5}) one gets
$$
p(\vec k, z) = p_0(\vec k) \frac{1 + \frac{\mu}{\sqrt{z^2 + (kv)^2}}}{z + Dk^2 + \lambda }
$$
Similar to the three dimensional case one has to distinguish between the limiting cases (i) $kv/z \gg 1$ and (ii) 
$kv/z \ll 1$. Whereas in the first case the initial and the final state $p \approx p_i + p_f$ 
is reflected, in the second case the solution can be approximately written as 
$$
p(\vec k, z) = \frac{p_s(\vec k}{z} + \Psi (\vec k, z)\,.
$$
As before for $v \to 0$ the system tends to a stationary solution of the form 
\begin{equation}
p_s(\vec x) = \frac{\mu }{ 2 \pi D} \int d^2x^{\prime} p_0(\vec x^{\prime}) 
K_0\left(\frac{\mid \vec x - \vec x^{\prime} \mid}{\xi }\right)\,.
\end{equation} 
We omit the further details. Instead of that let us consider the $d$-dimensional case. 
Because of the factor $1/z$ in front of the hypergeometric function in Eq.~(\ref{eq5}) we expect the occurrence of 
a pseudo-stationary for arbitrary $d$. For $t \gg [D k^2 + \lambda ]^{-1}$ Eq.~(\ref{eq3}) can be solved in leading order as 
\begin{equation}
p(\vec k ,t) \simeq p_0(\vec k) \lambda A \frac{J_{\frac{d-2}{2}}(kv t)}{(kvt)^{(d-2)/2} [D k^2 + \lambda ]}\,.
\label{eq9}
\end{equation}
This solution can be transformed in the coordinate representation $p(\vec x, t)$. The result is for 
$0< vt < r \equiv  \mid \vec x - \vec x' \mid$ 
\begin{equation}
p(\vec x, t) = \frac{\lambda \Gamma (d/2)}{D (vt)^{(d-2)/2} 2\pi ^{d/2}} 
\int d^dx^{\prime}\frac{p_0(\vec x^{\prime})}{(\mid \vec x - \vec x^{\prime} \mid)^{(d-2)/2}}
I_{\frac{d-2}{2}}(vt/ \xi) K_{\frac{d-2}{2}}(r/\xi )\,. 
\label{eq10}
\end{equation}
In the opposite case $0< r < vt$ we find  
\begin{equation}
p(\vec x, t) = \frac{\lambda \Gamma (d/2)}{D (vt)^{(d-2)/2} 2\pi ^{d/2}} 
\int d^dx^{\prime}\frac{p_0(\vec x^{\prime})}{(\mid \vec x - \vec x^{\prime} 
\mid)^{(d-2)/2}}I_{\frac{d-2}{2}}(r/ \xi) K_{\frac{d-2}{2}}(vt/\xi )\,. 
\label{eq11}
\end{equation}
Here $I_{\nu }(x)$ and $K_{\nu }(x)$ are Bessel functions. 
In $d=1$ one gets for $t \gg [D k^2 + \lambda ]^{-1}$ the stationary solution discussed in Eq.~(\ref{d13}). 

\section{Conclusions}

In this paper we have extended the conventional diffusion reaction system by including non-Markovian 
memory terms. In particular the growth rate is modified accordingly. Namely the particles are permanently 
subjected to signals of the environment, which acts like a background radiation. In case a particles is 
annihilated a recreation process sets in which is triggered by this background signals from the past. 
Alternatively the model can be viewed as a system of traps, distributed randomly at the initial time $t=0$. 
The particles immersed in the traps will be activated and released from the traps only when a particle is 
annihilated. Due to the competition between both processes one observes different time regimes with 
a different behavior. In particular the system evolves a transient regime with a very slowly varying 
concentration field which tends under special circumstances to a stationary pattern solution in 
according to previous studies. Such patterns are absent for conventional diffusion in an infinite 
domain without boundary conditions. 

\begin{acknowledgments} 
This work is supported by the DFG (SFB 418). 
\end{acknowledgments}

\end{document}